\documentclass{jpp}
\usepackage{amssymb}
\usepackage{amsmath}
\usepackage{graphicx}
\usepackage{bm}
\usepackage{mathrsfs}
\usepackage{cite}
\usepackage[latin1]{inputenc}

\newcommand{\be}{\begin{equation}}
\newcommand{\ee}{\end{equation}}
\newcommand{\ber}{\begin{eqnarray}}
\newcommand{\eer}{\end{eqnarray}}

\def\case#1/#2{\textstyle\frac{#1}{#2} }

\begin{document}

\title[Interaction between gravitational waves and plasma waves]{Interaction between gravitational waves and plasma waves in the
Vlasov description}
\author{G.\ns  B\ls R\ls O\ls D\ls I\ls N, M.\ns  F\ls O\ls R\ls S\ls B\ls E\ls R\ls G, M.\ns  M\ls A\ls R\ls K\ls L\ls U\ls N\ls D\ns  and\ns  D.\ns   E\ls R\ls I\ls K\ls S\ls S\ls O\ls N}
\affiliation{Department of Physics, Ume{\aa} University, SE--901 87 Ume{\aa}, Sweden}

\maketitle

\begin{abstract}
The nonlinear interaction between electromagnetic, electrostatic and
gravitational waves in a Vlasov plasma is reconsidered. By using a
orthonormal tetrad description the three-wave coupling coefficients are
computed. Comparing with previous results, it is found that the present
theory leads to algebraic expression that are much reduced, as compared to
those computed using a coordinate frame formalism. Furthermore, here we
calculate the back-reaction on the gravitational waves, and a simple energy
conservation law is deduced in the limit of a cold plasma. \\[2mm]
\end{abstract}


\section{Introduction}

Much work has been devoted to special relativistic effects of plasmas \cite%
{Shukla1986}, largely stimulated by the rapid progress in high power laser
technology \cite{MM+PK}. Several studies of plasmas in a general
relativistic context have also been made, see .e.g. \cite{%
Tajima1997,Scripta99,grishchuk,Papadouplous2001,Moortgat2003,Moortgat2004},
considering for example plasmas in strongly curved space-times close to
pulsars \cite{Membrane}, multi-fluid plasma effects in general relativity \cite{Marklund-etal2003,Betschart-etal2004,Rahman2009}, dynamo effects in strong gravity \cite{Marklund-Clarkson}, the effect of two-temperature systems on scalar perturbations \cite{moortgat-marklund2006}, viscous heating of accretion discs due to gravitational wave dissipation \cite{Kocsis}, or considering gravitational wave (GW) propagation in a plasma
medium, in particular nonlinear interactions in e.g. dusty plasmas \cite{Brodin-etal2005,Forsberg-etal2006} or MHD plasmas \cite{Kallberg2004,Duez1,Duez2,Isliker,Farris}. GW:s are currently opening up a promising new window for astronomy and astrophysics, e.g. astroseismology \cite{Andersson-Kokkotas,Samuelsson-etal2009}, and the interaction between GW:s and electromagnetic fields have been proposed as a possible means to detect GW:s \cite{Braginskii,Pegoraro,Caves,Brodin-Marklund2003,Stedman}. Moreover, the nonlinear interaction of a curved spacetime with and electromagnetic field can yield a multitude of interesting astrophysical and cosmological effects \cite{Tsagas2005,giovannini2006,Tsagas-Barrow-Maartens,Fenu-Durrer2009}. In a magnetized plasma, a basic effect is the linear coupling
between electromagnetic (EM) waves and GW:s that occurs for propagation
across a static magnetic field \cite{grishchuk}.\footnote{For a discussion concerning the 
case of cosmological magnetic fields, see Refs.\ \cite{Tsagas2003,Betschart2005,Zunkel2006,Tsagas2007,Betschart2007}.}
This linear mechanism leads
to the excitation of magnetohydrodynamic waves in a plasma \cite{%
Papadouplous2001,Moortgat2003,Moortgat2004}. Naturally linear coupling
mechanisms is not sufficient if one is interested in the possibility of
converting GW energy to frequencies different from that of the original
source. Numerous examples of such mechanisms exists in plasmas, involving
e.g. frequency up-conversion due to nonlinear wave steepening \cite{%
Forsberg2008},or various three-wave couplings between GWs and
electromagnetic waves, e.g. \cite{Brodin-Marklund1999,Servin2000}. Wave
coupling mechanisms involving GWs are studied for several different reasons.
In some cases, the emphasis is on the basic theory \cite{%
servinbrodin,Mendonca2002,Balakin2003,BMD2000}. In other works, the focus
is on GW detectors \cite{Li2002,BrodinMarklund2003,Picasso2003}, on
cosmology \cite{Papadoupolus2002,MDB2000,Hogan2002}, or on astrophysical
applications such as binary mergers \cite{bms}, gamma ray bursts \cite{%
Vlahos2004}, pulsars \cite{Mosquera2002} or supernovas \cite{Brodin2005}.

In the present paper we will re-consider the problem of three-wave
interacting between electrostatic (ES), electromagnetic (EM) and GW:s in a
plasma, using a collisionfree kinetic description, i.e. the Vlasov equation 
\cite{Brodin-Marklund1999}. In contrast to previous authors we will use a
tetrad formalism \cite{Cargese}, rather than a coordinate frame formalism,
since the former formalism has been shown to significantly reduce the
algebraic complexity. In particular, the coordinate frame formalism applied
on the present problem produced very cumbersome algebraic expressions for
the coupling coefficients \cite{Brodin-Marklund1999}, and as a result it was
impossible to see if the growth rates for parametric processes where
positive in general (when damping was omitted), which is related to the
fulfillment of the so called Manley-Rowe relations \cite{Weiland}. The
coefficients derived using the present tetrad formalism is shown to agree
with previous works in the limit of a cold plasma \cite{Brodin-Marklund1999}%
. For the case of a finite temperature, the coefficients found from the
present formalism are algebraically much simpler than previous results.
However, due to the complexity of the previously computed coefficients, a
comparison cannot be easily made for the general case. Finally, in the
present paper we also include the back-reaction on the GW, in contrast to
previous works. This allows us to discuss the energy conservation
properties, and an energy conservation law is presented in the low
temperature limit.

\section{Basic equations in the tetrad formalism}

We consider the interaction between weak gravitational waves and a
collisionless plasma in an external magnetic field. Since we consider
nonempty space the background space-time is necessarily curved. However, if
the wavelength of the gravitational waves and the interaction region is
small relative to the background curvature we may take the background to be
flat and static (the lowest order version of the high-frequency
approximation), and consider the perturbed energy-momentum tensor
corresponding to the \textit{perturbations} of the electromagnetic and
material fields.

For simplicity, the unperturbed plasma is assumed to be static, isotropic
and homogeneous. Linearized, the Einstein field equations (EFE) take the
form 
\begin{equation}
\square h_{ab}=-2\kappa \left[ \delta T_{ab}-{\tfrac{1}{2}}\delta T\eta
_{ab}\right]  \label{EFE}
\end{equation}%
provided the gauge condition $h_{\quad ,b}^{ab}=0$ is fulfilled, which is
equivalent to state that only tensorial perturbations are present. Here $%
\square \ \equiv \left[ c^{-2}\partial _{t}^{2}-\partial _{z}^{2}\right] $, $%
h_{ab}$ is the small deviation from the Minkowski background metric, i.e. $%
g_{ab}=\eta _{ab}+h_{ab}$, $\kappa \equiv 8\pi G/c^{4}$, $\delta T_{ab}$ is
the part of the energy-momentum tensor containing small electromagnetic and
material field perturbations associated with the gravitational waves and $%
\delta T$ $=\delta T_{\;a}^{a}$. In the following it is understood that we
neglect contributions of second order and higher in $h_{ab}$. In our
notations $a,b,c,...=0,1,2,3$ and $i,j,k,...=1,2,3$ and the metric has the
signature $(-+++)$.

In vacuum, a linearized gravitational wave can be transformed into the
transverse and traceless (TT) gauge. Then we have the following line-element
and corresponding orthonormal frame basis 
\begin{eqnarray}
\mathrm{d}s^{2} &=&-c^{2}\mathrm{d}t^{2}+(1+h_{+}(\xi ))\,\mathrm{d}%
x^{2}+(1-h_{+}(\xi ))\,\mathrm{d}y^{2}  \notag \\
&&+2h_{\times }(\xi )\,\mathrm{d}x\,\mathrm{d}y+\mathrm{d}z^{2}\ ,
\label{le}
\end{eqnarray}%
\begin{eqnarray}
\mathbf{e}_{0} &\equiv &c^{-1}\partial _{t}\ ,\quad \mathbf{e}_{1}\equiv (1-{%
\tfrac{1}{2}}h_{+})\partial _{x}-{\tfrac{1}{2}}h_{\times }\partial _{y}\ ,
\label{tetrad-1} \\
\mathbf{e}_{2} &\equiv &(1+{\tfrac{1}{2}}h_{+})\partial _{y}-{\tfrac{1}{2}}%
h_{\times }\partial _{x}\ ,\quad \mathbf{e}_{3}\equiv \partial _{z}\ .
\label{tetrad2}
\end{eqnarray}%
where $\xi \equiv z-ct$ and $h_{+},h_{\times }\ll 1$. As it turns out, the
gravitational waves takes this form also in the particular case (propagation
in an isotropic plasma) that we are considering. The difference to the
vacuum case will be that $\xi =z-v_{\mathrm{ph}}t$, where $v_{\mathrm{ph}}$
is the phase velocity of the gravitational wave. Note, however that the
theory will be limited to the case $v_{\mathrm{ph}}\approx c$, due to the
omittance of back-ground curvature effects. From now on we will refer to
tetrad components rather than coordinate components.

We follow the covariant approach presented in \cite{Cargese}] for splitting the electromagnetic and material fields in a $1+3$ fashion. Suppose an
observer moves with 4-velocity $u^{a}$. This observer will measure the
electric and magnetic fields $E_{a}\equiv F_{ab}u^{b}$ and $B_{a}\equiv 
\tfrac{1}{2}\,\epsilon _{abc}F^{bc}$ , respectively, where $F_{ab}$ is the
electromagnetic field tensor and $\epsilon _{abc}$ is the volume element on
hyper-surfaces orthogonal to $u^{a}$. It is convenient to introduce a
3-vector notation $\mathbf{E}\equiv (E^{i})=(E^{1},E^{2},E^{3})$ etc. and $%
\mathbf{\nabla }\equiv \mathbf{e}_{i}$. From now on we will assume that $%
u^{0}=c$ is the only nonzero component of $u^{a}$. As has been presented in
e.g. Refs.\ \cite{Marklund-Brodin-Dunsby2000,Servin2001}] the Maxwell equations contain terms coupling the
electromagnetic field to the gravitational radiation field. Including terms
that are linear in $h_{+}$ and $h_{\times }$, but omitting terms that are
quadratic and of higher order, Maxwell's equations are written as 
\begin{eqnarray}
\nabla \times \mathbf{B} &=&\mu _{0}(\mathbf{j+j}_{E})+\frac{1}{c^{2}}%
\partial _{t}\mathbf{E}, \\
\partial _{t}\mathbf{B} &=&\nabla \times \mathbf{E}  -\mathbf{j}%
_{B}, \\
\nabla \cdot \mathbf{B} &=&0  \label{divB} \\
\nabla \cdot \mathbf{E} &=&\frac{\rho _{c}}{\varepsilon _{0}}  \label{divE}
\end{eqnarray}%
where 
\begin{eqnarray}
\mathbf{j}_{E} &&=-\frac{1}{2c}\left[ (E_{x}-cB_{y})\dot{h}%
_{+}+(E_{y}+cB_{x})\dot{h}_{\times }\right] \mathbf{e}_{1}
\nonumber \\ && \quad
+\frac{1}{2}\left[
(E_{y}+cB_{x})\dot{h}_{+}+(E_{x}-cB_{y})\dot{h}_{\times }\right] \mathbf{e}%
_{2},  \label{JE-source} \\[2mm]
\mathbf{j}_{B} &&=-\frac{1}{2}\left[ (E_{y}+cB_{x})\dot{h}_{+}-(E_{x}-cB_{y})%
\dot{h}_{\times }\right] \mathbf{e}_{1}
\nonumber \\ && \quad
-\frac{1}{2}\left[ (E_{x}-cB_{y})\dot{%
h}_{+}+(E_{y}+cB_{x})\dot{h}_{\times }\right] \mathbf{e}_{2},
\label{JB-source}
\end{eqnarray}%
are effective currents due to the GWs, see e.g. Ref. \cite{BMD2000}], and the
dot denotes derivatives with respect to the argument. The physical current-
and charge-density are denoted $\mathbf{j}$ and $\rho _{c}$, respectively.
Note that the absence of a GW source terms in Eqs. (\ref{divB}) and (\ref%
{divE}) is valid only within the given approximation, as can be seen from
e.g. Ref. \cite{BMD2000}], where source terms that are fully nonlinear in the
GW-amplitude are included. In addition to the explicit source terms in (\ref%
{JE-source})-(\ref{JB-source}), naturally the gravitational effects that is
associated with the tetrad (\ref{tetrad-1})-(\ref{tetrad2}) must also be
kept in mind.

Next we turn our attention to the particle description. The equation of
motion for a particle of mass $m$ and charge $q$ in an electromagnetic and
gravitational wave field is 
\begin{equation}
\frac{d}{dt}\mathbf{p}=q\left[ \mathbf{E}+{(}\gamma m)^{-1}\mathbf{p{\times B%
}}\right] -\mathbf{G}  \label{eqmot}
\end{equation}%
where $\gamma =\sqrt{1+p_{i}p^{i}/(mc)^{2}}$ and the four-momenta is $%
p^{a}=\gamma mdx^{a}/dt$. The gravitational force like term $G^{i}\equiv
\Gamma _{\;ab}^{i}p^{a}p^{b}/\gamma m$, where $\Gamma _{\;ab}^{i}$ are the
Ricci rotation coefficients, becomes 
\begin{eqnarray}
G_{1} &=&{\tfrac{1}{2}}(v_{\mathrm{ph}}-p_{z}/\gamma m)\left[ \dot{h}%
_{+}p_{1}+\dot{h}_{\times }p_{2}\right]  \label{G1} \\
G_{2} &=&{\tfrac{1}{2}}(v_{\mathrm{ph}}-p_{z}/\gamma m)\left[ -\dot{h}%
_{+}p_{2}+\dot{h}_{\times }p_{1}\right]  \label{G2} \\
G_{3} &=&{\tfrac{1}{2}(}\gamma m)^{-1}\left[ \dot{h}%
_{+}(p_{1}^{2}-p_{2}^{2})+2\dot{h}_{\times }p_{1}p_{2}\right]  \label{G3}
\end{eqnarray}%
for weak gravitational waves propagating in the z-direction in Minkowski
space, where $v_{\mathrm{ph}}$ is the phase velocity of the gravitational
wave, which we at this point allow to deviate slightly from $c$.

Next we apply kinetic plasma theory, representing each particle species by a
distribution function $f$ governed by the Vlasov equation. In tetrad form
the Vlasov equation reads \cite{Servin2001}]
\begin{equation*}
\mathcal{L}f=0
\end{equation*}%
where the Liouville operator is 
\begin{equation*}
\mathcal{L}\equiv \partial _{t}+(c/p^{0})p^{i}e_{i}+\left[ F_{\mathrm{EM}%
}^{i}-\Gamma _{\;ab}^{i}p^{a}p^{b}c/p^{0}\right] \partial _{p^{i}}
\end{equation*}%
and the electromagnetic force responsible for geodesic deviation is $F_{%
\mathrm{EM}}^{i}\equiv q(E^{i}+\epsilon ^{ijk}p_{j}B_{k}/\gamma m)$. In
vector notation the Vlasov equation reads 
\begin{equation}
\partial _{t}f+\frac{\mathbf{p}\cdot \nabla f}{\gamma m}+\left[ q\left( 
\mathbf{E+}\frac{\mathbf{p\times B}}{\gamma m}\right) -\mathbf{G}\right]
\cdot \nabla _{\mathbf{p}}f=0  \label{Vlasov-vec}
\end{equation}%
where $\nabla _{\mathbf{p}}\equiv (\partial _{p_{1}},\partial
_{p_{2}},\partial _{p_{3}})$. In the absence of gravitational waves, the
Vlasov equation has the following spatially homogeneous (thermodynamical)
equilibrium solution, the Synge-J\"uttner distribution, e.g. \cite{%
Brodin-Marklund1999}, 
\begin{equation}
f_{SJ}=\frac{n_{0}\mu }{4\pi (mc)^{3}K_{2}(\mu )}e^{-\mu \gamma }  \label{SJ}
\end{equation}%
where $n_{0}$ is the spatial particle number density, $\mu \equiv
mc^{2}/k_{B}T$, $k_{B}$ is the Boltzmann constant, $T$ the temperature and $%
K_{2}(\mu )$ is a modified Bessel function of second kind.

\section{Wave-wave interaction}

Next we let all quantities consist of a superposition of three waves of
different kinds. Firstly we have a gravitational wave (frequency and
wavevector $(\omega _{\mathrm{g}},\mathbf{k}_{\mathrm{g}})$), next an
electromagnetic wave $(\omega _{\mathrm{em}},\mathbf{k}_{\mathrm{em}})$, and
finally an electrostatic wave (Langmuir wave) $(\omega _{\mathrm{es}},%
\mathbf{k}_{\mathrm{es}})$. Since we are dealing with high-frequency waves,
the mass and charge will in what follows refer to electrons. The three waves
are assumed to obey the following matching conditions%
\begin{equation}
\omega _{\mathrm{g}}=\omega _{\mathrm{em}}+\omega _{\mathrm{es}}
\label{Freq-energy}
\end{equation}%
and

\begin{equation}
\mathbf{k}_{\mathrm{g}}=k_{\mathrm{g}}\mathbf{e}_{3}=\mathbf{k}_{\mathrm{em}%
}+\mathbf{k}_{\mathrm{es}}  \label{Wave-vect-moment}
\end{equation}%
Since there is no external magnetic field within our model, the
gravitational wave does not induce an electric field at the frequency and
wavevector $(\omega _{g},\mathbf{k}_{g})$. Furthermore, the metric
perturbations at the combinations $(\omega _{1,2},\mathbf{k}_{1,2})$ can be
neglected, as a consequence of the high-frequency approximation.

As a prerequisite to the nonlinear calculations we first consider linear
theory. Linearizing the Vlasov equation, in the absence of EM-fields, the
perturbed distribution function of the GW is given by. 
\begin{equation}
f_{\mathrm{g}}=\frac{2ip_{a}G_{a}}{\hat{\omega}_{\mathrm{g}}}\frac{\partial
f_{SJ}}{\partial |p|^{2}}  \label{Gw-perturbed}
\end{equation}%
where we have introduced the notation $\hat{\omega}_{\mathrm{g}}=\omega _{%
\mathrm{g}}-k_{\mathrm{g}}cp_{3}/p_{0}$. Noting that the different
GW-polarizations obey 
\begin{equation}
\left( \omega _{\mathrm{g}}^{2}-k_{\mathrm{g}}^{2}c^{2}\right) h_{\times
}=-2\kappa \delta T_{12}=-2\kappa \int \frac{p_{1}p_{2}}{m\gamma }f_{\mathrm{%
g}}d^{3}p  \label{hx-1}
\end{equation}%
and%
\begin{equation}
\left( \omega _{\mathrm{g}}^{2}-k_{\mathrm{g}}^{2}c^{2}\right)
h_{+}=-2\kappa (\delta T_{11}-\delta T_{22})=-2\kappa \int \frac{%
(p_{1}^{2}-p_{2}^{2})}{m\gamma }f_{\mathrm{g}}d^{3}p  \label{h+1}
\end{equation}%
we deduce the same dispersion relation for both GW-polarizations, namely%
\begin{equation}
D_{g}(\omega _{\mathrm{g}},k_{\mathrm{g}})=\omega _{\mathrm{g}}^{2}-k_{%
\mathrm{g}}^{2}c^{2}-2\kappa n_{0}\int \frac{p_{1}^{2}p_{2}^{2}}{m\gamma }%
\frac{\partial f_{SJ}}{\partial |p|^{2}}d^{3}p=0  \label{DR-GW}
\end{equation}%
where we here and from now on use the normalization $\int f_{SJ}d^{3}p=1$ of
the unperturbed distribution. In the high-frequency approximation, the last
term of (\ref{DR-GW}) is a small correction, comparable to the contribution
from the background curvature \cite{grishchuk}], and thus we may use the the
approximation $D_{g}=\omega _{\mathrm{g}}^{2}-k_{\mathrm{g}}^{2}c^{2}$.
Similarly, from the linearized Vlasov equation (without gravitational
fields), together with Maxwell's equations we deduce the dispersion relation
for electromagnetic waves%
\begin{equation}
D_{\mathrm{em}}(\omega _{\mathrm{em}},k_{\mathrm{em}})=1-\frac{k_{\mathrm{em}%
}^{2}c^{2}}{\omega _{\mathrm{em}}^{2}}+\frac{\omega _{p}^{2}}{\omega _{%
\mathrm{em}}}\int \frac{2p_{2}^{2}}{\gamma \hat{\omega}_{\mathrm{em}}}\frac{%
\partial f_{SJ}}{\partial |p|^{2}}d^{3}p=0  \label{DR-EM}
\end{equation}%
and for electrostatic waves%
\begin{equation}
D_{\mathrm{es}}(\omega _{\mathrm{es}},k_{\mathrm{es}})=1-\frac{2m\omega
_{p}^{2}}{k_{\mathrm{es}}^{2}}\int \frac{\mathbf{k}_{\mathrm{es}}\cdot 
\mathbf{p}}{\gamma \hat{\omega}_{\mathrm{es}}}\frac{\partial f_{SJ}}{%
\partial |p|^{2}}d^{3}p=0  \label{DR-ES}
\end{equation}%
where $\omega _{p}=(n_{0}q^{2}/\varepsilon _{0}m)^{1/2}$ is the electron
plasma frequency. When nonlinear interactions are taken into account, the
wave amplitudes will be time-dependent. We note that as far as the linear
terms are concerned the only modification needed is the simple substitution $%
D_{\mathrm{es}}E_{\mathrm{es}}=$ $(\partial D_{\mathrm{es}}/\partial \omega
_{\mathrm{es}})\partial \tilde{E}_{\mathrm{es}}/\partial t$, where the tilde
denotes the weakly time-dependent amplitude, and similarly for the other
waves \cite{Weiland}]. Next, for definiteness, we assume the wave-vectors to
span the plane perpendicular to $\mathbf{e}_{2}$. Furthermore, we note that
for symmetry reasons, the $h_{\times }\,$-polarization in this geometry
couples to the EM-wave polarized with the electric field along $\mathbf{e}%
_{2}$ (nonlinearly combined with the electrostatic wave), whereas the $%
h_{+}\,$-polarization couples to the EM-waves with magnetic field along $%
\mathbf{e}_{2}$. These two cases are similar, and from now on we limit
ourselves to the former choice of polarization. The amplitude evolution for $%
\tilde{h}_{\times }$ is found by keeping resonant terms proportional to both
the EM and electrostatic wave in the energy-momentum tensor. Next we use Eq.
(\ref{hx-1}), generalized to keep the nonlinear terms and a weakly
time-dependent amplitude. Substituting the expression for $f_{g}$ from the
Vlasov equation, but now with second order nonlinear terms included, we
obtain 
\begin{equation}
\omega _{\mathrm{g}}\frac{\partial \tilde{h}_{\times }}{\partial t}%
=2\varepsilon _{0}\kappa C_{\mathrm{g}}\tilde{E}_{\mathrm{es}}\tilde{E}_{%
\mathrm{em}}  \label{coupling1}
\end{equation}%
where the coupling coefficient $C_{\mathrm{g}}$ is 
\begin{eqnarray}
C_{\mathrm{g}} &&=\left[ \frac{\mathbf{k}_{\mathrm{es}}\cdot \mathbf{e}_{1}}{k_{%
\mathrm{es}}}+\frac{\omega _{p}^{2}}{k_{\mathrm{es}}}\int \frac{%
p_{1}p_{2}^{2}}{\gamma \hat{\omega}_{\mathrm{em}}\hat{\omega}_{\mathrm{es}}}%
\right. \nonumber \\ && \quad \left. \times
\left\{ 2\frac{\mathbf{k}_{\mathrm{em}}\cdot \mathbf{k}_{\mathrm{es}}}{\hat{%
\omega}_{\mathrm{g}}\gamma m}\left[ \frac{\hat{\omega}_{\mathrm{em}}\omega _{%
\mathrm{es}}}{\omega _{\mathrm{em}}\hat{\omega}_{\mathrm{es}}}+\frac{\hat{%
\omega}_{\mathrm{es}}}{\hat{\omega}_{\mathrm{em}}}\right] \frac{\partial
f_{SJ}}{\partial \left\vert p\right\vert ^{2}}+4\mathbf{p}\cdot \mathbf{k}_{%
\mathrm{es}}\frac{\partial ^{2}f_{SJ}}{\partial (\left\vert p\right\vert
^{2})^{2}}\right\} d^{3}p.\right]   \label{coefficient1}
\end{eqnarray}%
To find the EM-wave evolution, we now include all resonant nonlinear source
terms for the EM-wave (effective gravitational currents, nonlinear terms
involving the gravitational force, and nonlinearities coming directly from
the tetrad), and solve for $\partial \tilde{E}_{\mathrm{em}}/\partial t$, in
which case we obtain 
\begin{equation}
\frac{\partial D_{\mathrm{em}}}{\partial \omega _{\mathrm{em}}}\frac{%
\partial \tilde{E}_{\mathrm{em}}}{\partial t}=-C_{\mathrm{em}}\tilde{h}%
_{\times }\tilde{E}_{\mathrm{es}}^{\ast }  \label{coupling2}
\end{equation}%
where 
\begin{eqnarray}
C_{\mathrm{em}} && =\left( -\frac{k_{\mathrm{em}}^{2}c^{2}}{2\omega _{\mathrm{em}%
}^{2}}-\frac{\omega _{\mathrm{g}}}{2\omega _{\mathrm{em}}}\right) \frac{%
\mathbf{k}_{es}\cdot \mathbf{e}_{1}}{k_{es}}  
+ \frac{\omega _{p}^{2}}{%
\omega _{\mathrm{em}}k_{es}}\int \frac{p_{2}^{2}}{\gamma \hat{\omega}_{%
\mathrm{em}}\hat{\omega}_{\mathrm{es}}}
\nonumber \\ &&\!\!\!\!\!\!\!\!\!\!\!\!\!\!\!\! \times
\left\{ \mathbf{k}_{\mathrm{es}}\cdot 
\mathbf{e}_{1}\left( 2\frac{\omega _{\mathrm{g}}\hat{\omega}_{\mathrm{es}}}{%
\hat{\omega}_{\mathrm{g}}}-\frac{\hat{\omega}_{\mathrm{g}}\omega _{\mathrm{es%
}}}{\hat{\omega}_{\mathrm{es}}}\right) \frac{\partial f_{SJ}}{\partial
\left\vert p\right\vert ^{2}}
-\frac{4p_{1}\omega _{\mathrm{g}}\hat{\omega}_{\mathrm{em}}}{\hat{%
\omega}_{\mathrm{g}}}\mathbf{p}\cdot \mathbf{k}_{\mathrm{es}}\frac{\partial
^{2}f_{SJ}}{\partial (\left\vert p\right\vert ^{2})^{2}}\right\} d^{3}p .
\label{Coefficient2b}
\end{eqnarray}%
Finally the calculation is completed by doing analog calculation for the
electrostatic wave evolution, and the result is 
\begin{equation}
\frac{\partial D_{\mathrm{es}}}{\partial \omega _{\mathrm{es}}}\frac{%
\partial \tilde{E}_{\mathrm{es}}}{\partial t}=-C_{\mathrm{es}}\tilde{h}%
_{\times }\tilde{E}_{\mathrm{em}}^{\ast }  \label{Coupling3}
\end{equation}%
where%
\begin{eqnarray}
&& C_{\mathrm{es}} =\frac{\omega _{p}^{2}}{k_{\mathrm{es}}}\left[ \frac{\mathbf{k%
}_{\mathrm{em}}\cdot \mathbf{e}_{1}}{2\omega _{p}^{2}}- \int \frac{p_{2}^{2}}{%
\gamma \hat{\omega}_{\mathrm{es}}}
\left( \frac{\mathbf{k}_{\mathrm{em}%
}\cdot \mathbf{e}_{1}}{\hat{\omega}_{\mathrm{em}}} - \frac{2\mathbf{k}_{%
\mathrm{es}}\cdot \mathbf{k}_{\mathrm{em}}\omega _{\mathrm{g}}p_{1}}{\gamma 
\hat{\omega}_{\mathrm{es}}\hat{\omega}_{\mathrm{g}}\omega _{\mathrm{em}}m_{e}%
} 
\right.\right. \nonumber \\[2mm] && \left.\left.\quad 
- \frac{ \left( \omega _{\mathrm{g}}- {p_{3}k_{\mathrm{g}}}/{\gamma
m_{e}}\right) \mathbf{k}_{\mathrm{es}}\cdot \mathbf{e}_{1}+ ({2p_{1}}/{%
\gamma m_{e}})\mathbf{k}_{\mathrm{es}}\cdot \mathbf{k}_{\mathrm{g}} }{%
\hat{\omega}_{\mathrm{em}}\hat{\omega}_{\mathrm{es}}}\right) \frac{\partial
f_{SJ}}{\partial |p|^{2}}d^{3}p\right]   \label{Coefficient-3}
\end{eqnarray}%
The coupling coefficients $C_{\mathrm{es}}$ and $C_{\mathrm{em}}$ (although
not $C_{\mathrm{g}}$) has been calculated in Ref. \cite{Brodin-Marklund1999}]%
, using a coordinate frame formalism. As can be seen the previously computed
coefficients are algebraically much more complicated, and thus a comparison
is difficult in general. For the special case where the temperature of $%
f_{SJ}$ approaches zero, the coefficients simplify a lot, and we obtain%
\begin{equation}
C_{\mathrm{es}}=C_{\mathrm{em}}=\frac{\mathbf{k}_{\mathrm{es}}\cdot \mathbf{e%
}_{1}\omega _{\mathrm{g}}}{k_{\mathrm{es}}\omega _{\mathrm{em}}}
\label{Comparison}
\end{equation}%
which agrees with Ref. \cite{Brodin-Marklund1999}]. Furthermore, we find that
in our case that also the gravitational case coefficient is similar, and we
can define the common coupling coefficient $C\equiv C_{\mathrm{es}}=C_{%
\mathrm{em}}=C_{\mathrm{g}}$. As a result of the agreement of the
coefficients, the energy change of each of the waves in the cold limit can
be written%
\begin{eqnarray}
	&& \frac{dW_{\mathrm{g}}}{dt} 		= -\omega _{\mathrm{g}}V,  \label{E1} \\
	&& \frac{dW_{\mathrm{es}}}{dt} 	=\omega _{\mathrm{es}}V  \label{E2}
\end{eqnarray}%
and 
\begin{equation}
\frac{dW_{\mathrm{em}}}{dt}=\omega _{\mathrm{em}}V  \label{E3}
\end{equation}%
where the energy density of each wave is 
\begin{equation}
	W_{\mathrm{g}}=\omega _{\mathrm{g}%
}^{2}\vert \tilde{h}_{\times }\vert ^{2}/2\kappa , 
\end{equation}
\begin{equation}
	W_{\mathrm{es}}=\varepsilon _{0}\omega _{\mathrm{es}}(\partial D_{\mathrm{es}}/\partial
\omega _{\mathrm{es}})\vert \tilde{E}_{\mathrm{es}}\vert ^{2} ,
\end{equation}
\begin{equation}
	W_{\mathrm{em}}=\varepsilon _{0}\omega _{\mathrm{em}}(\partial D_{%
\mathrm{em}}/\partial \omega _{\mathrm{em}})\vert \tilde{E}_{\mathrm{em}%
}\vert ^{2}
\end{equation}
and%
\begin{equation}
V=\varepsilon _{0}C\tilde{h}_{\times }\tilde{E}_{\mathrm{em}}^{\ast }\tilde{E%
}_{\mathrm{es}}^{\ast }+\mathrm{c.c}  \label{V-coeff}
\end{equation}%
where $\mathrm{c.c.}$ denotes complex conjugate. The equations (\ref{E1})-(\ref{E3})
together with (\ref{Freq-energy}) thus shows that the three-wave interaction
process conserves the total wave-energy.

\section{Summary and conclusion}

In the present paper we have reconsidered the process of three-wave
interaction between gravitational-, electromagnetic- and electrostatic waves
in a collisionfree plasma. Using an ortonormal tetrad description, the
algebraic complexity of the coupling coefficients is much reduced, as
compared to previous authors \cite{Brodin-Marklund1999}]. Two of the
coupling coefficients (for the electrostatic and the electromagnetic wave)
can be shown to agree with previous results in the cold limit. The third
coefficient (for the GW) has not been calculated before. Inclusion of the
back-reaction on the GW makes it possible to deduce and energy conservation
law (Eqs (\ref{E1})-(\ref{E3}) together with (\ref{Freq-energy})) in the
cold limit. The energy conservation law is a natural consequence of the
Manley-Rowe relations \cite{Weiland}], which makes the same coefficient
appear in all three coupled equations (\textit{c.f.}\ $C\equiv C_{\mathrm{es}}=C_{%
\mathrm{em}}=C_{\mathrm{g}}$.). However, this relation was only verified in
the cold limit. Although the present coefficients were not extremely
complicated, it has proved to be difficult to deduce whether the coupling
coefficients are symmetric (i.e. fulfilling the Manley-Rowe relations), for
the general case of a finite temperature. An interesting question, that
remains for future research, is thus to decide whether this is due to some
principal lack of a canonical Hamiltonian structure (that is the basic
source of symmetric coupling coefficients, see e.g. \cite{Larsson-JPP}]) of
the Einstein-Maxwell-Vlasov system, or if the difficulties are merely due to
the algebraical complexity.

\acknowledgments This research was partially supported by the Swedish
Research Council and the Swedish Graduate School of Space Technology.

\end{document}